# Investigation of radiation-enhanced diffusion using first-passage time

V. V. Ryazanov

Institute for Nuclear Research, Kiev, Ukraine, e-mail: vryazan19@gmail.com

Highlights
- Diffusion time is considered first-passage time
- Obtained expressions for the diffusion coefficient
- Theoretical results are compared with experimental ones

The approach the first-passage time (*FPT*) of a random process to a certain level is applied to the description of radiation-enhanced diffusion. This is an integral approach to describing the problem of radiation-enhanced diffusion, which does not specify the details of the process but takes into account the overall change in entropy during the process. Relationships are obtained to take into account the influence of radiation effects on the diffusion coefficient. The theoretical results obtained are compared with the experimental ones.

## 1. Introduction

Radiation-enhanced diffusion (*RED*) is a fundamental process that determines the evolution of the microstructure in materials under radiation exposure. For real structural materials of nuclear power plants, including austenitic chromium-nickel steels, *RED* is a limiting process of radiation damage: radiation-induced segregation, pore formation, and swelling, effects of radiation creep and growth, etc. [1-17].

For polycrystalline materials with impurities, the *RED* process is enhanced by the contribution of radiation-stimulated boundary diffusion and the formation of rapidly diffusing complexes of impurity atoms with point defects at grain boundaries. The phenomenon of radiation-enhanced diffusion has been known for a long time and was studied in many works [1-23].

The role of irradiation is manifested in the intensification of diffusion processes due to the formation of point defects that reduce the activation energy of diffusion, in the forced movement of atoms in the crystal lattice due to atom-atom collisions (recoil atoms), in a change in the elemental composition due to nuclear reactions. The formation of point radiation defects in irradiated materials, and primarily vacancies, transfer the material as a thermodynamic system to a new state, more or less stable relative to the existing metastable state.

Under irradiation, radiation-accelerated processes occur, for example, decomposition of solid solutions, ordering, growth of nuclei of the second phase, and radiation-induced processes, for example, separation and segregation of elements, formation of a second phase, shape change (through dissolution) of particles of the second phase, disordering, which are based on are diffusion processes.

In this paper, an integral approach is used to describe the problem of *RED*, which does not specify the details of the process, but takes into account the overall change in entropy during the process. In the physical process of reaching a certain boundary by a diffusing particle, a change in entropy occurs, which must be taken into account. This approach is based on using the first-passage time (*FPT*) of a random process to a certain level [24, 25].

*FPT* proved to be a powerful way to describe systems of various natures. In [26], the *FPT* is described, and the relationship between the *FPT* and the change in entropy in the system is considered. Diffusion processes were also considered [27]. In one example (drift-diffusion process), additional changes in entropy, caused, among other things, by radiative influences, lead only to a decrease in *FPT*, which is consistent with accelerated diffusion, as *RED* is sometimes called. However, another example, with Feller processes, diffusion processes with constant drift and diffusion coefficients, shows that entropy changes are possible, in which the average first-passage time to reach the level can increase [26, 27]. This situation corresponds not to acceleration, but to deceleration of diffusion. Such phenomena are also observed, for example, in semiconductors [28].

The use of *FPT* in this work is based on the determination of the diffusion coefficient *D*, (22)-(23), when $D = \frac{(b-Y_0)^2}{\bar{T}_\gamma} = D_0 \frac{T_0}{\bar{T}_\gamma}$, $D_0 = \frac{(b-Y_0)^2}{T_0}$, where $\bar{T}_\gamma$, $T_0$ are the average *FPTs* for reaching the boundary by a diffusing particle in the presence and absence of radiative effects, *b* and 0 are boundaries of the diffusion region, $Y_0=Y(t=0)$ is the initial position of the diffusing particle, $D_0$ is the thermal diffusion coefficient without taking into account radiation effects, *D* is the diffusion coefficient taking into account radiation effects. The bar above $\bar{T}_\gamma$ denotes the mean value of the random variable $T_\gamma$, $T_0 = \bar{T}_{\gamma=0}$, and the subscript *γ* indicates the dependence of this value on the parameter *γ* from the distribution (1), which relates $\bar{T}_\gamma$ to the change in entropy during the *FPT* time, and distinguishes $T_\gamma$ from the temperature $T_1$ from (1), *T* (14).

There are also processes of anomalous diffusion, which are not considered in this paper. In general, diffusion, despite its long history of research, is still not a fully understood phenomenon. This is due to the variety of diffusion mechanisms and a very large number of heterogeneous materials with different structures in which diffusion occurs. Therefore, an attempt to find general laws that are valid for many substances, media, and physical processes of diffusion, which is undertaken in this work, may be useful.

Results similar to those of the present work for inhomogeneous transport and inference from first-passage statistics were obtained in [29, 30]. But the method of taking into account the effects on the system in [29, 30] differs from the method used in this work. The approach of this work is based on the results of [26]. Another similar approach to the problem under consideration and also different in the way of taking into account the impacts on the process is carried out in [31]. In a simplified form, the relationship between *FPT* and entropy for a particular case of diffusion processes was considered in [32].

Section 2 briefly outlines the general approach to [26], in which the average value of *FPT* is associated with the change in entropy over the *FPT* time. In Section 3, an explicit stochastic model of the process under consideration is given. Section 4 compares calculations based on the theoretical model with experimental results. In conclusion, brief conclusions are presented.

## 2. The statistical description of the first-passage time and entropy change

In all non-equilibrium processes, entropy changes occur. This also applies to *FPT* processes, during which the entropy of the system changes. The relationship between *FPT* and entropy changes during *FPT* is considered in [26]. In [26, 33-36], a generalization of the Gibbs distribution to the nonequilibrium case is introduced, containing *FPT* as a thermodynamic parameter. The introduction of such a distribution is based on the theory of large deviations and the maximum entropy method. In this case, a statistical distribution is used, which contains the time it takes for a random process to reach a certain level (first-passage time [34, 35] or lifetime in terms of [36]).

In [26, 33-36], the random value of the first passage time (lifetime) $T_\gamma$ of a random process was considered as a thermodynamic parameter. A statistical distribution with this parameter was introduced. The microscopic density of this distribution in the phase space of variables $z$ ($z=(q_1,...,q_N; p_1,...,p_N)$ for a system of $N$ particles, coordinates $q_i$ and momenta $p_i$ of all particles of the system) has the form

$$\rho(z;u,T_\gamma) = \exp\{-\beta u - \gamma T_\gamma\}/Z(\beta,\gamma), \qquad \beta = 1/T_1, \qquad (1)$$

where $u$ is the energy of the system (considered as a random process), $T_1$ is the temperature (in energy units), $\gamma$ is the thermodynamic parameter conjugate to the random value of the first passage time (lifetime) $T_\gamma$, similar to how the reverse temperature $\beta = 1/T_1$ is conjugate to random energy $u$; can be called parameter $\gamma$ a tempentropy;

$$Z(\beta,\gamma) = \int \exp\{-\beta u - \gamma T_\gamma\}dz = \int\int du dT_\gamma \omega(u,T_\gamma) \exp\{-\beta u - \gamma T_\gamma\} \qquad (2)$$

is the partition function. Equality in expression (2) corresponds to the transition from a microscopic description to a macroscopic one. The Lagrange parameters $\beta$ and $\gamma$ are determined from the equations for the averages:

$$\langle u \rangle = -\partial \ln Z/\partial \beta_{|\gamma}; \qquad \langle T_\gamma \rangle = -\partial \ln Z/\partial \gamma_{|\beta}. \qquad (3)$$

In [26, 33-36] cells of the "extended" (compared to the Gibbsian) phase space with constant values $(u, T_\gamma)$ are introduced (instead of phase cells with constant values of $u$). The structure factor $\omega(u)$ is replaced by $\omega(u,T_\gamma)$ - the volume of the hypersurface in the phase space containing fixed values of $u$ and $T_\gamma$. In this case $\int \omega(u,T_\gamma)dT_\gamma = \omega(u)$. The function $\omega(u,T_\gamma)$ corresponds to the internal properties of the system. At the mesoscopic level, this function is described by a random process, and $\omega(u,T_\gamma)$ is the joint probability density of unperturbed (for $\gamma=0$ and $\beta = \beta_{eq}$ is an equilibrium state (or some stationary nonequilibrium) and without external influences) values $u$, $T_\gamma$ understood as stationary distribution of this process. Distribution (1)-(2) generalizes the Gibbs distribution, in which $\gamma=0$, and is valid for stationary non-equilibrium systems. In [37] it is shown that non-stationary processes can also be described by this kind of distribution.

The joint distribution density of random variables $u$ and $T_\gamma$ is equal to

$$p(u,T_\gamma) = \int \delta(u - u(z))\delta(T_\gamma - T_\gamma(z))\rho(z;u(z)T_\gamma(z))dz =$$
$$= \rho(u,T_\gamma)\omega(u,T_\gamma) = \exp\{-\beta u - \gamma T_\gamma\}\omega(u,T_\gamma)/Z(\beta,\gamma). \qquad (4)$$

Integrating (4) over $T_\gamma$, we obtain a distribution for the energy $u$ of the form

$$p(u) = \int p(u, T_\gamma) dT_\gamma = \frac{e^{-u\beta}}{Z(\beta,\gamma)} \int_0^\infty \omega(u, T_\gamma) e^{-\gamma T_\gamma} dT_\gamma, \qquad \beta = 1/T_1. \qquad (5)$$

Following the assumptions about the form of the function $\omega(u, T_\gamma)$, we rewrite it in the form

$$\omega(u, T_\gamma) = \omega(u)\omega_1(u, T_\gamma) = \omega(u) \sum_{i=0}^n P_i f_i(T_\gamma, u). \qquad (6)$$

Expression (6) assumes that the system has $n+1$ classes of ergodic states, $P_i$ is the probability that the system is in the $i$-th class of ergodic states, $f_i(T_\gamma, u)$ is the probability density that in a system that is in ergodic states of the $i$-th class, the time of the first passage is $T_\gamma$. The expression under the integral on the right side of (5) determines the non-equilibrium part of the distribution.

If we choose the function $f_i$ in (6), for example, in the form of a gamma distribution

$$f_i(x) = \frac{1}{\Gamma(\alpha_i)} \frac{1}{b_i^{\alpha_i}} x^{\alpha_i - 1} e^{-x/b_i}, \quad x > 0, \quad f_i(x) = 0; \ x < 0; \quad \int_0^\infty e^{-\gamma_i x} f_i(x) dx = (1 + \gamma_i b_i)^{-\alpha_i}, \qquad (7)$$

where $\Gamma(\alpha)$ is the gamma function, $b_i$, $\alpha_i$ - distribution parameters, and put $b_i \alpha_i = \bar{T}_{\gamma 0 i}$, ($\bar{T}_{\gamma 0 i}$ are the unperturbed average lifetimes of the system in $i$-th states [26, 33-36]), $\alpha_i = \gamma_i / \lambda_i$, where $\lambda_i$ is the intensity of entry into the $i$-th subsystem in a state of dynamic equilibrium, $\gamma_i = \gamma$ in the $i$-th subsystem, then from (4)-(7) we obtain that

$$(1 + \gamma_i b_i)^{-\alpha_i} = (1 + \lambda_i T_{\gamma 0 i})^{-\gamma_i / \lambda_i}; \quad p(u) = \int p(u, T_\gamma) dT_\gamma = \frac{e^{-\beta u} \omega(u)}{Z(\beta, \gamma)} \sum_{i=0}^n P_i / (1 + \lambda_i T_{\gamma 0 i})^{\alpha_i}. \qquad (8)$$

If $f_i$ is an exponential distribution, then $\alpha_i = 1$, $\gamma_i = \lambda_i$, $b_i = \bar{T}_{\gamma 0 i}$. For the exponential distribution when $\alpha_i = 1$ in (7)-(8) and one class of ergodic states, when $n = 0$,

$$(1 + \gamma T_{\gamma = 0})^{-1} = <T_\gamma>/T_{\gamma = 0}, \qquad (9)$$

where $<T_\gamma> = \bar{T}_\gamma$ is the average lifetime of the system obtained from (2)-(3), (6)-(7) [33-36].

Let us define partition function (2). In this case, we use expressions (2)-(3) and the approximation that the distribution of the first passage time $f(T_\gamma)$ does not depend on the random value of energy, the variables are separated, and equilibrium and nonequilibrium components are distinguished in the total average energy [26],

$$\bar{u} = -\frac{\partial \ln Z}{\partial \beta}\bigg|_\gamma, \quad Z(\beta, \gamma) = Z_\beta Z_\gamma, \quad Z_\beta = \int e^{-\beta u} \omega(u) du, \quad Z_\gamma = \int_0^\infty e^{-\gamma T_\gamma} f(T_\gamma) dT_\gamma, \quad \beta = 1/T_1, \qquad (10)$$

$$Z_\gamma = \int_0^\infty e^{-\gamma T_\gamma} \sum_{j=0}^n P_j f_j(T_\gamma, u) dT_\gamma. \qquad (11)$$

where $f(T_\gamma, u) = f(T_\gamma, \bar{u}) = f(T_\gamma)$ is the probability density of the distribution of the *FPT*. In (10), we assume that the parameters included in the distribution of the first passage time depend on the average values of the energy, and not on their random values. By expressions (10)-(11), the partition function factor describing the non-equilibrium behavior of the system $Z_\gamma$ is expressed in terms of the Laplace transform of the first-passage time $T_\gamma$ distribution density $f(T_\gamma)$.

Besides distribution (7), many other distributions for *FPT* can be used. It depends on the specific task.

The parameter $\gamma$ of distribution (1) will be related to entropy using the generally accepted definition of entropy in statistical physics as the logarithm of the distribution density (1) averaged over this distribution, $s = \langle \ln \rho(z, u, T_\gamma) \rangle$, where brackets denote averaging. Variables are separated as in (10). In the case of one class of ergodic states, from (6) we obtain $\omega(u, T_\gamma) = \omega(u) f(T_\gamma)$, $\omega(T_\gamma) = f(T_\gamma)$. The distribution density (4) is equal to $p(u, T_\gamma) = \frac{e^{-\beta u} \omega(u)}{Z_\beta} \frac{e^{-\gamma T_\gamma} f(T_\gamma)}{Z_\gamma} = p(u) p(T_\gamma)$, where $Z_\beta$, $Z_\gamma$ defined in (10). The entropy of this distribution is $s = -k_B \int p(u, T_\gamma) \ln[p(u, T_\gamma)] du dT_\gamma = s_\beta + s_\gamma$, $s_\beta = \beta \bar{u} + \ln Z_\beta$, $s_\gamma = \gamma \bar{T}_\gamma + \ln Z_\gamma$. We took into account that for the nonequilibrium case in [37], the relation $s = s_B + \langle s(B) \rangle$, $s_B = -k_B \int p(B) \ln[p(B)] dB$, $\langle s(B) \rangle = k_B \int p(B) \ln[\omega(B)] dB$ was obtained; $B$ are random internal thermodynamic parameters, functions of dynamic variables $z$; in our case $B_1 = u$, $B_2 = T_\gamma$. The total uncertainty in the system is equal to the sum of the uncertainty of the parameters $B$ and the average uncertainty of the dynamic variables remaining after fixing the parameter $B$. Expanding in a power series in powers $\gamma$ value $s_\gamma$, we get $s_\gamma = -\gamma^2 (\langle T_0^2 \rangle - \langle T_0 \rangle^2) \leq 0$, $s \to s/k_B$, entropy is divided by $k_B$, Boltzmann's constant.

Thus, the terms $\int \frac{e^{-\beta u} \omega(u)}{Z_\beta} \ln[\omega(u)] du$, $\int \frac{e^{-\gamma T_\gamma} f(T_\gamma)}{Z_\gamma} \ln[f(T_\gamma)] dT_\gamma$ cancel and

$$s = s_\gamma + s_\beta = s_\beta - \Delta = \gamma \bar{T}_\gamma + \beta \bar{u} + \ln Z = \beta \bar{u} + \ln Z_\beta - \Delta, \quad -\Delta = s_\gamma = s - s_\beta. \quad (12)$$

($s$ is the entropy density). We got a match with the expression $s = \langle \ln \rho(z, u, T_\gamma) \rangle$.

Expression (13) has been refined in comparison with the similar expression from [26] and follows from (10)-(12) and the definition of entropy as the average logarithm of the distribution (1)

$$-\Delta = s_\gamma = \gamma \bar{T}_\gamma + \ln Z_\gamma. \quad (13)$$

where the quantities $Z_\gamma$, $\bar{T}_\gamma$, are defined in the relations (2)-(3), (10)-(11), $\Delta \geq 0$.

## 3. Stochastic process model

It is required to establish an adequate correspondence between a physical phenomenon and a random process used for its mathematical modeling. Many random processes can be adapted to a specific task. The properties of the process can be significantly affected by various circumstances, for example, changes in the boundary conditions [38, 39].

The same phenomenon under different conditions can exhibit different properties and be described by different random processes, distributions and their Laplace transforms. Examples of this kind are given in [40]. For our purposes, the Laplace transforms of the *FPT* distributions are more important than the distributions themselves. This facilitates the analytical solution of the problem.

The phenomenon under study is modeled by some random process. The functional of this random process is the *FPT* statistics. The exact analytical form of the solution for the *FPT* is determined in rare cases. It is possible to use general results for the *FPT*, for example, the first few terms of the expansions obtained in [41-42] for the Laplace transform of the *FPT*. There are other options for selecting and configuring *FPT* statistics models. The use of various types of approximations also significantly depends on the stage of system evolution [43-44].

The proposed approach is based on probabilistic mathematical relations and has great generality. The area of applicability of the results is large and, possibly, corresponds to the area of applicability of the Gibbs statistics. This generality allows one to consider arbitrary distributions rather than being restricted to any one particular form of *FPT*.

We are interested in the following Markov process of a random variable *Y* (for example [45])

$$\frac{dY(t)}{dt} = -D\frac{\partial \beta V(Y,t)}{\partial Y} + \sqrt{2D}\xi(t), \qquad (14)$$

where *Y* is the position of the diffusing particle, *D* is the diffusion constant and $\xi(t)$ satisfies the white noise condition $\langle \xi(t) \rangle = 0$, $\langle \xi(t_1)\xi(t_2) \rangle = 2D\delta(t_1 - t_2)$, brackets denote averaging. The time-dependent potential *V(Y,t)* has a discontinuity at a point *b*,

$$V(Y,t) = \begin{cases} V_0(Y,t), & Y \leq b \\ -\infty, & Y = b \end{cases}, \qquad (15)$$

where $\beta=1/k_BT$, $k_B$ is the Boltzmann constant, *T* is absolute temperature, and $V_0(Y,t)$ is a smooth function, differentiable for the range $-\infty < Y < \infty$ at an arbitrary time. This random process may correspond to the diffusion of a particle in an external field $V_0(Y)$ in the limit of large friction, but – in addition to this – the particle may disappear when passing over point *b*. Assuming that the particle is initially located at $Y_0=Y(t=0)$ ($0<Y_0<b$), $b>0$ with the probability of 1.

Process (14) is considered on the interval (0, *b*). In [38], a differential equation was obtained for the probability density of the first-passage time reaching the boundary *b*. After applying the Laplace transform to this equation, taking into account the initial condition, a second-order ordinary differential equation was obtained in [38] for the Laplace transform of the probability density of the first-passage time reaching the boundary. In our case, this is the value $Z_\gamma$ (11).

We will consider the case of the linear potential (15) $V_0(Y)=-FY$. In real physical systems, there are various forces *F* arising from intramolecular and intermolecular interactions. For example, the resistance force of the medium of a moving particle. Equation (14) can be rewritten in the form

$$\frac{dY(t)}{dt} = \mu + \sqrt{2D}\xi(t), \qquad (16)$$

where $\mu = \beta FD \sim D/b$, $\beta = 1/k_BT$. For the Laplace transform $Z_\gamma$ of the probability density of the *FPT* reaching the boundaries, in [38] an ordinary linear differential equation of the second order was obtained, the general solution of which for the case (16) has the form

$$Z_\gamma = Z_{(0,b)}(s_1 = \gamma, Y_0) = Ae^{\gamma_1 Y_0} + Be^{\gamma_2 Y_0}, \qquad (17)$$

where $s_1=\gamma$ from (11), $Y_0=Y(t=0)$ is the initial value of random variable *Y*. Constants *A* and *B* are determined from the boundary conditions. If both boundaries 0, and *b* are absorbing, then the boundary conditions have the form

$$Z_{(0,b)}(s_1,0) = Z_{(0,b)}(s_1,b) = 1, \quad Ae^{\gamma_1 b} + Be^{\gamma_2 b} = 1, \quad A+B = 1, \qquad (18)$$

and $\gamma_1$, $\gamma_2$ are roots of the characteristic equation

$$\gamma_{1,2} = \frac{-\mu \mp \sqrt{\mu^2 + 4Ds_1}}{2D}. \qquad (19)$$

From the boundary conditions (18) of the relation (17), we find the constants *A* and *B* and for the Laplace transform (11) we obtain the expression

$$Z_\gamma = \int_0^\infty e^{-\gamma t} h(t) dt = \frac{e^{\gamma_2 Y_0}(1-e^{\gamma_1 b}) - e^{\gamma_1 Y_0}(1-e^{\gamma_2 b})}{(e^{\gamma_2 b} - e^{\gamma_1 b})}, \qquad (20)$$

where the numerator and denominator in (20) are positive at $s_1>0$, $b>0$; $Z_{s_1}<1$.

In [46] three types of *FPT* self-similarity are introduced: (i) stochastic, which holds in 'real space'; (ii) Laplace, which holds in 'Laplace space'; and (iii) joint, which is the combination of the stochastic and Laplace types. Analysis establishes that the three types of *FPT* selfsimilarity yield, respectively and universally, the following *FPT* distributions: inverse-gamma; inverse-Gauss; and Levy–Smirnov. Moreover, the analysis explicitly pinpoints the classes of diffusion processes that produce the three types of self-similar *FPT*s. Shifting from general diffusion dynamics to Langevin dynamics, it is shown that the three classes collapse, respectively, to the following specific processes: diffusion in a logarithmic potential; Brownian motion with drift; and Brownian motion. We are reviewing Brownian motion with drift.

We will put $\beta F \sim 1/b$. We write the dependence on $\beta$ at $k_B = 8.6\times10^{-5}$ eVK$^{-1}$ and $x=10^4/T$ as $\beta = x/0.86$. Then $\beta F \sim x/\bar{x}b$, where $\bar{x}$ is the average value of the *x* parameter in the selected temperature range. The magnitude of the acting force *F* is generally unknown. Changes in the value of *F* will be specified by the parameter *k*, assuming $\beta F \sim 1/kb$. We assume the dependence of the diffusion coefficient *D* on the parameter $\beta$ in the form $D \sim \exp(-\beta E_a)$, $E_a$ is the activation energy. The force *F* does not depend on $\beta$, and $\partial F/\partial\beta = 0$.

The parameter $s_1=\gamma$, which is included in (19), is determined from equation (13). If from expression (20) with the help of relations (3), (10), and (20) we find the terms of equation (13), then we obtain a complex transcendental equation for $s_1=\gamma$. We use the series expansion in $s_1=\gamma$ of expressions (20) and expressions obtained from (20) for $\bar{T}_\gamma$. Equation (13) takes the form

$$-\Delta = \sum_{k=2}^\infty \frac{(k-1)}{k!} \frac{\partial^k(-\ln Z_{s_1})}{\partial s_1^k}\bigg|_{s_1=0}.$$ A quadratic expansion in $s_1=\gamma$ is used. This expression looks like

$$s_1 = \gamma = \left[\frac{2\Delta(1-e^{-1/k_1})}{A}\right]^{1/2}, \qquad (21)$$

$$A = \frac{b^4 k_1^2}{D^2}[(2k_1-1)(1-e^{-\alpha/k_1}) - 4\alpha e^{-\alpha/k_1} - 2k_1\alpha(1-e^{-1/k_1}) + \frac{(e^{-\alpha/k_1}-1)(e^{-\alpha/k_1}+1+2e^{-1/k_1})}{e^{-1/k_1}-1}], \quad k_1 = \frac{k}{x/\bar{x}}, \quad x=\frac{10^4}{T}.$$

The influence of radiation on the average *FPT* $\bar{T}_\gamma$ is included in the parameter $\Delta$ from (12)-(13), (21).

The diffusion coefficient, by its definition, is related to the average value of reaching the diffusion boundary by the relation

$$D = \frac{(b-Y_0)^2}{\bar{T}_\gamma} = D_0 \frac{T_0}{\bar{T}_\gamma} = D_0 y, \quad y = \frac{T_0}{\bar{T}_\gamma}, \quad D_0 = \frac{(b-Y_0)^2}{T_0}, \qquad (22)$$

where in the expression for *D*, the radiation effects are taken into account through the parameter $\Delta$ by expression (23); these effects are not taken into account in the expression for $D_0$.

For the value of *y* from (22), after performing the described calculations, we obtain the expression

$$T_0 = \frac{b^2 k_1}{D} \left( \frac{1 - e^{-\alpha/k_1}}{1 - e^{-1/k_1}} - \alpha \right), \quad Y_0 = \alpha b, \quad 0 < \alpha < 1, \quad k_1 = \frac{k}{x/\bar{x}}, \quad x = \frac{10^4}{T}, \quad \mu = D/bk_1, \qquad (23)$$

$$y = \frac{T_0}{\bar{T}_\gamma}, \quad \bar{T}_\gamma = \frac{b^2 k_1}{D\sqrt{1 + 4D\gamma/\mu^2}} \left[ \frac{\alpha[e^{\gamma_1 Y_0}(1 - e^{\gamma_2 b}) + e^{\gamma_2 Y_0}(1 - e^{\gamma_1 b})] + e^{\gamma_1 Y_0} e^{\gamma_2 b} + e^{\gamma_2 Y_0} e^{\gamma_1 b}}{e^{\gamma_1 Y_0}(1 - e^{\gamma_2 b}) - e^{\gamma_2 Y_0}(1 - e^{\gamma_1 b})} - \frac{e^{\gamma_2 b} + e^{\gamma_1 b}}{e^{\gamma_1 b} - e^{\gamma_2 b}} \right],$$

where $\bar{x}$ is the average value of $x$ in the range of the parameter $x$, the values $\gamma_1, \gamma_2$ from relation (19). In the values of $\gamma_1, \gamma_2$ (19), from (23) are substituted with the values $s_1 = \gamma$ (21), determined from (13) through the change in entropy, which includes radiative effects. The value of $y$ does not depend on the parameters $b$ and $D$, they are canceled in (21)-(23).

When calculating Fig.1-4 section 4, the parameter $\Delta$ was set from the results of [50] as the vacancy formation entropy $s^f/k_B = -s_\gamma = \Delta = 2$ ($\Delta$, $s_\gamma$ from (12)- (13)). The value of $\Delta$ one can also be determined from extended irreversible thermodynamics [52], as in [27] (in this case, the full Langevin equation for a massive particle should be used). There are other possibilities for determining the value of $\Delta$ [53].

Note that, in addition to process (14), it is possible to use both some other continuous random processes and some discrete random processes. Once again, we note that for each specific task, it is necessary to look for that random process that most fully corresponds to this particular task.

Equation (14) was also studied under various boundary conditions. The region (0, $b$) was considered. In the case of absorbing boundary conditions at point $b$ and reflecting at point 0, the non-equilibrium part of the partition function (10), which coincides with the Laplace transform of the *FPT* distribution, is equal to

$$Z_\gamma = \int_0^\infty e^{-\gamma t} h(t) dt = \frac{e^{\gamma_2 Y_0}(1 - (\gamma_2/\gamma_1) e^{(\gamma_1 - \gamma_2) Y_0})}{e^{\gamma_2 b}(1 - (\gamma_2/\gamma_1) e^{(\gamma_1 - \gamma_2) b})}, \quad \gamma_{1,2} = \frac{-\mu \mp \sqrt{\mu^2 + 4D\gamma}}{2D}, \qquad (24)$$

Similar results are also obtained using the Wald distribution [27], when $y = D/D_0 = D^{irrad}/D^{thermal} = T_0/\bar{T}_s = 1 + 2\Delta k_1/(1-\alpha) \pm \sqrt{(1 + 2\Delta k_1/(1-\alpha))^2 - 1}$. In this case, there are two branches of the solution. When choosing a sign – parameter $y<1$.

## 4. Comparison of calculations by expressions (21)-(23) with experimental results

Let us compare the calculations using the obtained expressions with the experimental results from [8]. We set the beginning of the diffusion region at zero and consider one-dimensional diffusion on the interval (0, $b$). The results depend on the initial values, the parameter $\alpha$, and the value of $k$ associated with the acting force $F$.

Qualitatively, the behavior of the diffusion coefficient in Figs. 1-4 is the same as in Figs. 4, 8 in [8]. The value of $D$ from (22) corresponds to the value of $D^{irrad}$ from [8], and the value of $D_0$ corresponds to the value $D^{thermal} = 0.48 \exp(-3.22/k_B T)$ from [8]. As in [8], the enhancement factor, $y = D^{irrad}/D^{thermal}$, was increased as the irradiation temperature decreased. There are quantitative differences. The main discrepancy lies in the dependence of the behavior of the enhancement factor $y$ (23) on the increase in the parameter $x$. However, for example, the differences between the experimental works [8] and [47] are much larger. The reason for this may be different values of the initial position $\alpha$ and acting forces $F(k)$ in the diffusion medium. Computer evaluations of the obtained results show that such dependencies have a very strong influence on the final results. It is possible that other stochastic models of diffusion, for example, various forms of the Feller process [48, 49], will lead to a better agreement with the experimental results. The use of the Feller process, as shown in [26, 27], can lead to a decrease in the diffusion coefficient, which is also observed experimentally for other temperature intervals. In [27], expressions for the concentration and diffusion fluxes were also obtained. Vacancy (interstitial) concentration induced by irradiation is $C_V^{irrad}$, $C_V^{thermal}$ is the vacancy concentration at thermal equilibrium condition and the value of $C_V^{thermal}$ was evaluated as follows: $C_V^{thermal} = \exp(S_f/k_B T)\exp(-E_V^f/k_B T)$, where $S^f$ is the entropy and $E_V^f$ is the vacancy formation energy; $S^f/k_B$ was set at 2, and $E_V^f$ was set to range from 1.4 [50] to 2.0 eV [51]. Perhaps, in the general case, other effects should also be taken into account.

The parameter $y$ (23) under (22) determines the deviation of the diffusion coefficient $D$ from the diffusion coefficient $D_0$, in which radiation effects are not taken into account

On fig. 4 [8] there is a point at which $D=D_0$, there is a transition from a decrease in the diffusion coefficient $D$ to its growth. There is no such point in the calculations using expressions (21)-(23). The diffusion coefficient $D$ is greater than the diffusion coefficient $D_0$ without taking into account radiation effects. The dependences of $D$ and $D_0$ on different $x$ scales and for different values of the parameters $α$ and $k$ are shown in Figs. 1-2. Figure 1 shows the behavior of the perturbed by radiation $D$ and unperturbed diffusion coefficients $D_0$ on the scale of Fig. 8 [8], at $x$: (11, 13), and Fig. 2 on the scale of $x$ from 8 to 24. Perhaps, with a weaker force $F$ acting from the medium, which corresponds to an increase in the parameter $k$, the situation will change. On fig. 3 shows the chaotic behavior of the enhancement factor $y = D^{irrad}/D^{thermal}$ (23) on the increase in the parameter $x$ at $k=500$. When $x \sim 10^6$ the enhancement factor $y \leq 1$. On fig. 4, 8 [8], a strong nonlinear increase in the diffusion coefficient $D$ compared to $D_0$ is also observed after crossing the point of equality between $D$ and $D_0$. In the results obtained in Fig. 1, 2 there is no such strong growth (as well as the intersection point itself). Similar results were also obtained for calculations according to equation (24) with an absorbing boundary at point $b$ and a reflecting boundary at point 0 (24) (in an equation of the form (21) for $s_1=γ$, the following expansion term must be taken into account), as well as for the Wald distribution [27]. The last result, close to Fig. 1, 2 for the Wald distribution and $D$ and $D_0$ is shown in Fig.4. Similar results were also obtained for the values $Δ>2$. At $Δ=5$, the agreement with the experiment is better than at $Δ=2$. This apparently shows that besides the vacancy formation entropy $s^f/k_B=-s_γ=Δ=2$ there are other entropy changes during the *FPT*.

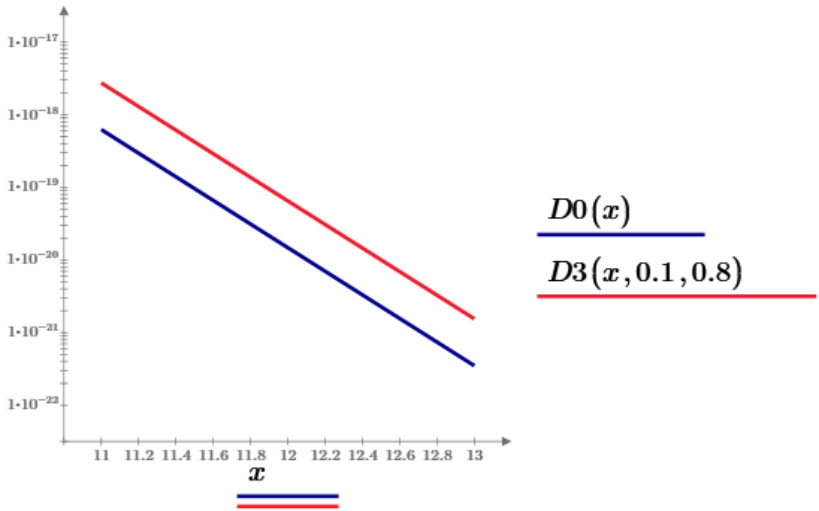

Fig.1. Dependences of $D = D^{irrad}$ (in red) and $D_0 = D^{irrad}$ (in blue) on the reciprocal temperature $x = 10^4/T$ in the interval $x$ (11, 13) as in Fig. 8 [8]. The initial values, the parameter $\alpha$ from $Y_0 = Y(t=0) = \alpha\, b$, and the value of $k$ associated with the acting force $F$, are set as $\alpha=0.1$, $k=0.8$. Values on the $y$-axis, as in Fig. 8 [8] are plotted on a logarithmic scale.

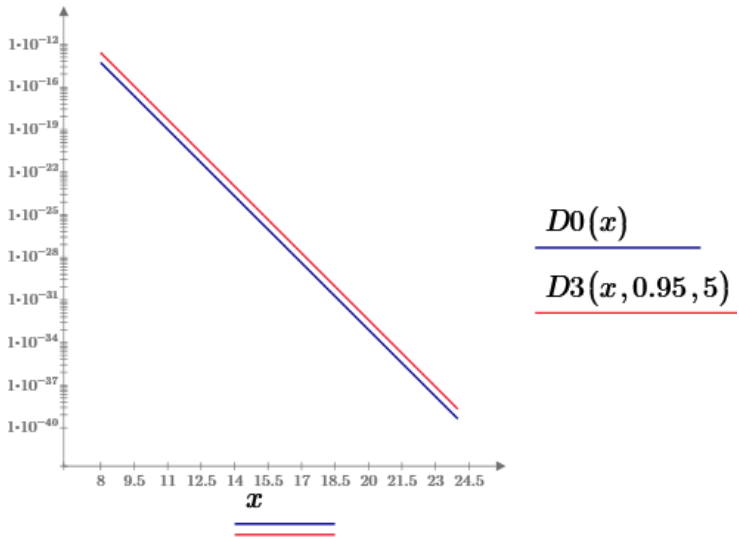

Fig.2. Dependences of $D = D^{irrad}$ (in red) and $D_0 = D^{irrad}$ (in blue) on the reciprocal temperature $x = 10^4/T$ in the interval $x$ (8, 24) as in Fig. 4 [8]. The initial values, the parameter $\alpha$ from $Y_0 = Y(t=0) = \alpha\, b$, and the value of $k$ associated with the acting force $F$, are set as $\alpha=0.95$, $k=5$. Values on the $y$-axis, as in Fig. 8 [8] are plotted on a logarithmic scale.

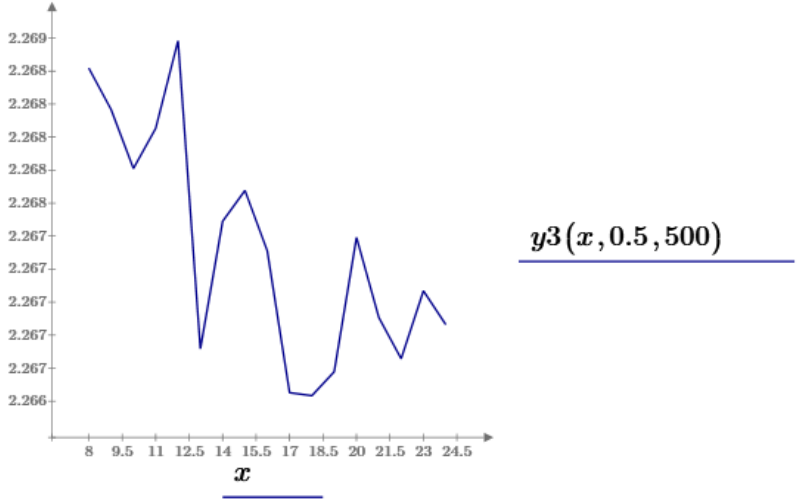

Fig.3. Dependence of the relation $y=D/D_0=D^{irrad}/D^{thermal}$ on the return temperature $x=10^4/T$ in the interval $x$ (8, 24). The initial values, the parameter $\alpha$ of $Y_0=Y(t=0)=\alpha b$, and the value of $k$ associated with the acting force $F$, are set as $\alpha=0.5$, $k=500$.

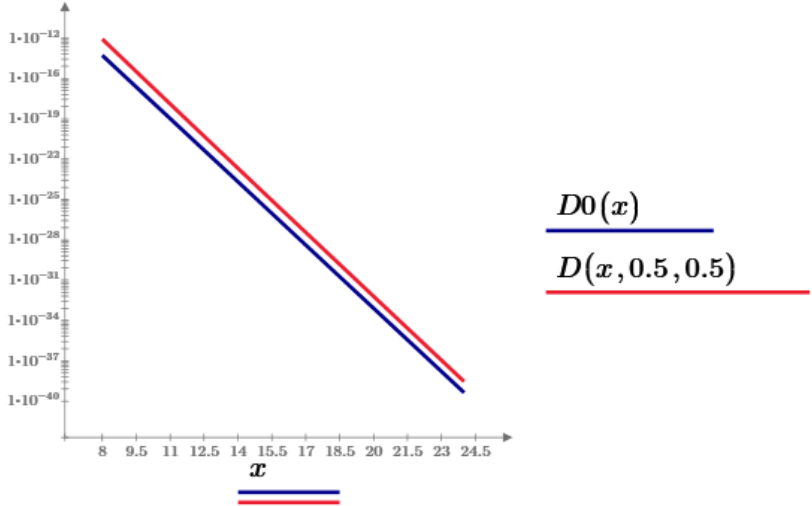

Fig.4. Dependences $D=D^{irrad}$ (in red) and $D_0=D^{thermal}$ (in blue) on the reciprocal temperature $x=10^4/T$ in the interval $x$ (8, 24), as in Fig. 8 [8]. The calculation was carried out in accordance with the Wald distribution [27]. The big decision branch with the + sign is used.

The obtained relations like (22)-(23) also help to determine the vacancy (interstitial) concentration induced by irradiation $C_V^{irrad}$. If the contribution of interstitials to diffusion is negligible, then [8] $D^{irrad}/D^{thermal}=(C_V^{thermal}+C_V^{irrad})/C_V^{thermal}$. Since the relation $D^{irrad}/D^{thermal}=y=(T_0/\bar{T}_\gamma)$ is defined in expression (23), the relation $C_V^{irrad}/C_V^{thermal}$ is also defined. Using a simple linear model [27] of the form $D(B)\approx D_0+BC(z)$, $B\sim\varphi$, where $\varphi$ is the radiation flux, in [27] an expression was written for the concentration $C(z)$ from the coordinate $z$, $l_0=Bc_0/D_0$, $C_0=C(z=0)$, $C_b=C(z=b)$, $C(z)=C_0[\sqrt{(1+1/l_0)^2-(1-C_b/C_0)(1+C_b/C_0+2/l_0)z/b}-1/l_0]$. These expressions are valid for $C_V^{irrad}$ and for $C_V^{thermal}$. We assume that the diffusion coefficient $D$ does not depend on the coordinate. In this case, the diffusion flux is equal to

$$J(z) = -D\frac{dC(z)}{dz} = D\frac{(1-C_b/C_0)(1+C_b/C_0+2/l_0)C_0/b}{2\sqrt{(1+1/l_0)^2 - (1-C_b/C_0)(1+C_b/C_0+2/l_0)z/b}}.$$

## 5. Conclusion

Diffusion is a complex phenomenon. Diffusion processes are diverse. For example, in solutions, chemical interactions and the formation of associates, complexes of molecules, significantly affect the processes of mutual diffusion. Even in the simple case of one-dimensional the drift-diffusion process of the form (14), for boundary conditions different from those considered in Section 3, it is possible to increase $\bar{T}_\gamma$ under the action of $\Delta$, rather than decrease, as for the Laplace transform (20) and Wald distribution [27]. Combinations of reflective and absorbing boundaries, partially absorbing boundaries, etc. are possible [38]. Significantly affect the solution and the initial position of the particles $\alpha$ and the forces $F(k)$ acting in the medium. The used stochastic model contains two unknown and arbitrary parameters. This is the initial position of the particle, characterized by the parameter $\alpha$, and the force $F(k)$ acting on the particle from the side of the medium. By changing the parameters $\alpha$ and $k$, a better agreement with the experiment can be obtained. Various patterns of diffusion processes are characteristic of various substances. There are many different mathematical models of diffusion processes. Although [46] explicitly identifies classes of diffusion processes leading to three types of self-similar *FPT*s: diffusion in a logarithmic potential; Brownian motion with drift; and Brownian motion.

It is possible to set and solve the problem of finding a certain mathematical model that most accurately corresponds to a specific diffusion process in a certain substance, under given physical conditions, both at the boundaries of the system and within it. Some mathematical models may not be implemented in real physical systems. *RED* is implemented only under certain conditions. Values $\Delta$ may not be reached at which accelerated diffusion can turn into slow diffusion.

During irradiation, the inhomogeneity of the types of defects that arise, and the variety of their interactions with each other and with the atoms of the substance create the possibility of sometimes directly opposite effects, for example, acceleration or deceleration of diffusion. In the model of the drift-diffusion process under consideration, when the system is affected, it is mainly possible to reduce the average *FTP*. This corresponds to the acceleration of diffusion. In the diffusion model of the Feller processes, the conditions and areas of influence are determined under which the average *FPT* can increase, which corresponds to a slowing down of diffusion [26, 27]. The detailed relationship between the obtained mathematical relationships and the physical mechanisms corresponding to these mathematical models should be investigated in detail. There are a large number of mathematical models of diffusion. Among recent papers, we can note the result obtained in [54], which generalizes the well-known result on the connection between the normal distribution and a parabolic differential equation of diffusion type to stable distributions and pseudo-parabolic equations corresponding to anomalous diffusion processes.

This paper provides a general overview of the application of the *FPT* method to *RED* studies. Diffusion by the vacancy (interstitial) mechanism is not considered. The advantage of the proposed approach to the description of *RED* is its generality. It applies not only to metals and alloys but to arbitrary physical systems. This consideration is illustrated by the example of the drift-diffusion process. The application of the described approach to a specific physical system requires setting an explicit stochastic model of the process. In this case, it is possible to use the results for estimating the transition through the energy barrier using the *FPT* study (for example, [55]), as well as other possibilities of the *FPT* method.

Note that the choice of the stochastic model is ambiguous. So, the boundary conditions, details, and features of the process can change. For example, in [56], the statistical properties of first-passage time functionals of a one-dimensional Brownian motion in the presence of stochastic resetting were studied.